\documentclass[twocolumn, showpacs, nofootinbib, aps, prd]{revtex4}

\usepackage{graphicx} \usepackage{dcolumn} \usepackage{color}
\usepackage{amsfonts}
\usepackage[colorlinks=true,linkcolor=blue,citecolor=blue,
urlcolor=blue]{hyperref}

\begin{document}

\title{The longitudinal and transverse distributions of the pion
wavefunction from the present experimental data on the pion-photon
transition form factor}

\author{Tao Zhong$^{1}$}
\email{zhongtao@htu.edu.cn}
\author{Xing-Gang Wu$^{2}$}
\email{wuxg@cqu.edu.cn}
\author{Tao Huang$^3$}
\email{huangtao@ihep.ac.cn}

\address{$^1$ Department of Physics, Henan Normal University, Xinxiang
453007, P.R. China\\ $^2$ Department of Physics, Chongqing University,
Chongqing 401331, P.R. China \\ $^3$ Institute of High Energy Physics and
Theoretical Physics Center for Science Facilities, Chinese Academy of
Sciences, Beijing 100049, P.R. China}

\date{\today}

\begin{abstract}

It is noted that the low-energy behavior of the pion-photon transition
form factor $F_{\pi\gamma}(Q^2)$ is sensitive to the transverse
distribution of the pion wavefunction, and its high-energy behavior is
sensitive to the longitudinal one. Thus a careful study on
$F_{\pi\gamma}(Q^2)$ can provide helpful information on the pion
wavefunction precisely. In this paper, we present a combined analysis of
the data on $F_{\pi\gamma}(Q^2)$ reported by the CELLO, the CLEO, the BABAR and the BELLE collaborations. It is performed by using the method of least
squares. By using the combined measurements of BELLE and CLEO
Collaborations, the pion wavefunction longitudinal and transverse behavior
can be fixed to a certain degree, i.e. we obtain $\beta \in [0.691,0.757]
\rm GeV$ and $B \in [0.00,0.235]$ for $P_{\chi^2} \geq 90\%$, where
$\beta$ and $B$ are two parameters of a convenient pion wavefunction model
whose distribution amplitude can mimic the various longitudinal behavior
under proper choice of parameters. We observe that the CELLO, CLEO and
BELLE data are consistent with each other, all of which prefers the
asymptotic-like distribution amplitude; while the BABAR data prefers a
more broad distribution amplitude, such as the CZ-like one.

\end{abstract}


\maketitle

\section{Introduction}

The pion-photon transition form factor (TFF) $F_{\pi\gamma}(Q^2)$ provides
a simplest example for the perturbative QCD (pQCD) application to
exclusive processes, where $Q^2$ stands for the momentum transfer. The TFF
relates two photons with the lightest meson (pion) and provides a good
platform for studying longitudinal and transverse properties of the pion
wavefunction.

The pion-photon TFF $F_{\pi\gamma}(Q^2)$ can be measured via the process,
$e^+ e^- \to e^+ e^- \pi^0$, in a single-tagged mode. It was first
measured in a low-energy region $Q^2 < 3 \rm GeV^2$ by the CELLO
collaboration~\cite{CELLO}. Later on, it was measured by the CLEO
collaboration~\cite{CLEO} in the energy region $Q^2 \in [1.5,9.2] \rm
GeV^2$, and by the BABAR collaboration~\cite{BABAR} and the BELLE
collaboration~\cite{BELLE} in the widest energy region $Q^2 \in [4,40] \rm
GeV^2$. On the other hand, it has been theoretically predicted by using
the pQCD approach, the QCD light-cone sum rules, or some phenomenological
models such as the semi-bosonized Nambu-Jona-Lasinio model and the
nonlocal chiral-quark model~\cite{TFF_AS, TFF_KT, TFF_KT96, TFF_KT97,
TFF_KT00, TFF_KT02, TFF_KT03, TFF_NV, TFF_NLO, TFF_BB122, TFF_BB131,
TFF_BB12, TFF_BB142, TFF_BB134, TFF_BB14, TFF_Rev15}. For example, Lepage
and Brodsky studied the pion-photon TFF by neglecting the transverse
distributions ($\textbf{k}_\perp$-distribution) of the constitute quarks,
and resulted in the well-known asymptotic prediction~\cite{TFF_AS}, i.e.,
$Q^2 F_{\pi\gamma}(Q^2)$ tends to be a constant ($\sqrt{2}f_\pi$) for the
asymptotic pion DA $\phi^{\rm as}_\pi(x,Q^2)|_{Q^2 \to \infty} = 6x(1-x)$.
The pion decay constant $f_\pi = 130.41 \pm 0.03 \pm 0.20 \rm
MeV$~\cite{PDG}.

When $Q^2$ $\sim$ a few $\rm GeV^2$, one should take into account the
$\textbf{k}_\perp$-corrections such that to achieve a reliable prediction
of $F_{\pi\gamma}(Q^2)$~\cite{TFF_KT, TFF_KT96, TFF_KT97, TFF_KT00,
TFF_KT02, TFF_KT03, TFF_NV}. The experimental data in this $Q^2$-region is
then helpful for determining the transverse behavior of the pion
wavefunction. When $Q^2$ is large enough, the $\textbf{k}_\perp$-terms
become less important, and the pion-photon TFF shall be dominated by the
longitudinal behavior of the pion wavefunction,which is related to the
pion distribution amplitude (DA). At present, there is no definite
conclusion on the pion DA due to the dramatic difference between the BABAR
and BELLE data. The experimental data in the large $Q^2$-region is thus
helpful for determining the longitudinal behavior of the pion
wavefunction.

In the paper, we shall study the pion-photon TFF $F_{\pi\gamma}(Q^2)$ by
using a convenient pion wavefunction constructed from the revised
light-cone harmonic oscillator model, whose DA can conveniently mimic the
Asymptotic-like to more broad longitudinal behavior via proper choices of
input parameters. Then we perform a combined analysis of the experimental
data reported by the CELLO, the CLEO, the BABAR and the BELLE
collaborations, with an attempt to extract useful information of the pion
wavefunction. For the purpose, we shall adopt the analytical expression of
$F_{\pi\gamma}(Q^2)$ suggested in our previous paper~\cite{TFF_NLO} as its
basic fitting function. The pion wavefunction parameters shall then be
fitted by comparing the experimental data with the help of the method of
least squares such that to achieve the best goodness-of-fit.

The remaining parts of the paper are organized as follows. In Sec.II, we
give a short review on the pion-photon TFF $F_{\pi\gamma}(Q^2)$ and a
brief introduction of the method of least squares. A combined analysis for
the experimental data reported by the CELLO, the CLEO, the BABAR and the
BELLE collaborations is presented in Sec.III. In Sec.IV, we analyze the
TFF $F_{\pi\gamma}(Q^2)$ in detail by using the BELLE and the CLEO data as
an attempt to find more accurate information on the pion wavefunction.
Sec.V is reserved for a summary.

\section{A Brief Review of the Pion-Photon TFF and the Method of Least
Squares}

The pion-photon TFF can be divided into two parts \begin{eqnarray}
F_{\pi\gamma}(Q^2) = F_{\pi\gamma}^{(V)}(Q^2) + F_{\pi\gamma}^{(NV)}(Q^2).
\label{pptff} \end{eqnarray} $F_{\pi\gamma}^{(V)}(Q^2)$ stands for the
contribution from the valence-quark part, which is pQCD calculable. The
analytical expression of $F_{\pi\gamma}^{(V)}(Q^2)$ can be found in
Ref.\cite{TFF_NLO}, in which the next-to-leading order
contributions~\cite{NLO97, NLO07, NLO09} and the
$\textbf{k}_\perp$-dependence has been kept explicitly, i.e.
\begin{widetext} \begin{eqnarray} F^{(V)}_{\pi \gamma}(Q^2)&=&
\frac{1}{4\sqrt{3}\pi^2}\int_0^1\int_0^{x^2 Q^2}\frac{dx}{x
Q^2}\left[1-\frac{C_F
\alpha_s(Q^2)}{4\pi}\left(\ln\frac{\mu_f^2}{xQ^2+k_\perp^2} +2\ln{x}+3-
\frac{\pi^2}{3} \right)\right] \cdot \Psi_{q\bar{q}}(x,k_\perp^2) d
k^2_\perp , \end{eqnarray} \end{widetext} where
$[dx]=dxdx'\delta(1-x-x')$, $C_F=4/3$ and $k_\perp=|\mathbf{k}_\perp|$.
$\mu_f=Q$ stands for the factorization scale. $F_{\pi\gamma}^{(NV)}(Q^2)$
stands for the nonvalence-quark part contribution that is related to the
higher Fock states of pion, which can be estimated via a proper
phenomenological model~\cite{TFF_NV}, \begin{equation} F^{(NV)}_{\pi
\gamma}(Q^2)=\frac{\alpha}{(1+Q^2/\kappa^2)^2} , \end{equation} where
$\kappa=\sqrt{-\frac{F_{\pi\gamma}(0)}{\frac{\partial}{\partial Q^2}
F^{(NV)}_{\pi \gamma}(Q^2)|_{Q^2\to 0}}}$ and
$\alpha=\frac{1}{2}F_{\pi\gamma}(0)$. It indicates $F^{(NV)}_{\pi
\gamma}(Q^2)$ is $1/Q^2$-suppressed to $F^{(V)}_{\pi \gamma}(Q^2)$ in
large $Q^2$-region, then it gives negligible contribution to the TFF at
large $Q^2$-region.

The pion-photon TFF $F_{\pi\gamma}(Q^2)$ is a convolution of hard
scattering amplitude with the $\textbf{k}_\perp$-correction and the pion
wavefunction. By taking the BHL-prescription~\cite{BHL}, the pion
wavefunction can be constructed over the light-cone harmonic oscillator
model~\cite{TFF_NLO}, i.e. \begin{eqnarray}
\Psi_{q\bar{q}}(x,\textbf{k}_\perp) = \frac{m_q}{\sqrt{\textbf{k}_\perp^2
+ m_q^2}} A \varphi(x) \exp \left[ -\frac{\textbf{k}^2_\perp +
m_q^2}{8\beta^2 x(1-x)} \right], \label{pion_T2WF} \end{eqnarray} where
$m_q$ is the mass of constituent quark, $A$ is the normalization constant,
$\beta$ is the harmonic parameter, and $\varphi(x) = 1 + B \times
C^{3/2}_2(2x-1)$ dominates the longitudinal distribution with the
Gegenbauer polynomial $C^{3/2}_2(2x-1)$. After integrating over the
transverse momentum dependence, we obtain the pion DA \begin{widetext}
\begin{eqnarray} \phi_{q\bar{q}}(x,\mu^2_0) &=& \frac{\sqrt{3}A m_q
\beta}{2\pi^{3/2}f_\pi} \sqrt{x(1-x)} \varphi(x) \times \left\{ {\rm Erf}
\left[ \sqrt{\frac{m_q^2 + \mu_0^2}{8\beta^2 x(1-x)}} \right] - {\rm Erf}
\left[ \sqrt{\frac{m_q^2}{8\beta^2 x(1-x)}} \right] \right\},
\label{pion_T2DA} \end{eqnarray} \end{widetext} where the initial scale
$\mu_0 \sim 1$ GeV, the error function ${\rm Erf}(x) =
\frac{2}{\sqrt{\pi}} \int^x_0 d^{-t^2} dt$. The pion DA satisfies the
normalization condition, $\int^1_0 dx \phi_{q\bar{q}}(x,\mu^2_0) = 1$. The
input model parameters can be fitted from the known experimental data. In
addition, one extra constraint from the sum rules of $\pi^0 \to
\gamma\gamma$ shall be adopted, which states $\int^1_0 dx
\Psi_{q\bar{q}}(x,\textbf{k}_\perp = 0) = \frac{\sqrt{6}}{f_\pi}$. We can
adopt this and the normalization condition to fix the values of $A$ and
$\beta$, leaving $m_q$ and $B$ as the two free parameters to be determined
from the data. Ref.\cite{TFF_NLO} has observed that if setting $m_q$ to be
the usually choosing 300 MeV, the pion DA (\ref{pion_T2DA}) shall change
from the asymptotic-like form~\cite{TFF_AS} to the CZ-like
form~\cite{CZ_DA} by simply shifting the parameter $B$ from $0.00$ to
$0.60$. In the present paper, to be more general, we shall adopt a broader
range $m_q \in [200, 400]\rm MeV$ and $B \in [0.00,0.60]$ to do our fit.

The data fit shall be done by using the method of least squares.
Considering a set of $N$ independent measurements $y_i$ with the known
variance $\sigma_i$ and the mean $\mu(x_i;\mathbf{\theta})$ at known
points $x_i$. The measurements $y_i$ are assumed to be in Gaussian
distribution. The goal of the method of least squares is to get the
preferable value of $\mathbf{\theta}$ by minimizing the likelihood
function~\cite{PDG} \begin{eqnarray} \chi^2(\mathbf{\theta}) =
\sum^N_{i=1} \frac{(y_i - \mu(x_i,\mathbf{\theta}))^2}{\sigma_i^2}.
\label{ls} \end{eqnarray} As for the present case, the function
$\mu(x_i;\mathbf{\theta})$ stands for the pion-photon TFF function defined
by (\ref{pptff}) and $\mathbf{\theta} = (m_q, B)$; The value of $y_i$ and
its variance $\sigma_i$ for the pion-photon TFF can be read from the
measurements of the CELLO, the CLEO, the BABAR and the BELLE
collaborations~\cite{CELLO, CLEO, BABAR, BELLE}, respectively. The
goodness-of-fit is judged by the magnitude of the probability
\begin{eqnarray} P_{\chi^2} = \int^\infty_{\chi^2} f(y;n_d) dy,
\label{px2} \end{eqnarray} where $f(y;n_d) =
\frac{1}{\Gamma(\frac{n_d}{2}) 2^{n_d/2}} y^{\frac{n_d}{2}-1}
e^{-\frac{y}{2}}$ is the probability density function of $\chi^2$, and
$n_d$ is the number of degree-of-freedom. The probability $P_{\chi^2}$ is
within the range of $[0,1]$; when its value is closer to $1$, a better fit
is assumed to be achieved.

\section{Best fit of the CELLO, the CLEO, the BABAR and the BELLE data on
$F_{\pi\gamma}(Q^2)$}

\begin{table}[htb] \begin{tabular}{ c | c c c c } \hline ~~ & ~CELLO~ &
~CLEO~ & ~BABAR~ & ~BELLE~ \\ \hline ~$m_q(\rm MeV)$        ~ & ~$216
$~ & ~$246     $~ & ~$347      $~ & ~$222     $~ \\ ~$B$ ~
~ & ~$0.000  $~ & ~$0.002   $~ & ~$0.600    $~ & ~$0.000   $~ \\ ~$A(\rm
GeV^{-1})$     ~ & ~$20.695 $~ & ~$21.823  $~ & ~$19.438   $~ & ~$20.906
$~ \\ ~$\beta(\rm GeV)$      ~ & ~$0.801  $~ & ~$0.697   $~ & ~$0.664
$~ & ~$0.776   $~ \\ ~$\chi^2_{\rm min}/n_d$~ & ~$4.795/3$~ & ~$4.380/13$~
& ~$15.508/15$~ & ~$5.657/13$~ \\ ~$P_{\chi^2_{\rm min}}$~ & ~$0.187  $~ &
~$0.986   $~ & ~$0.416    $~ & ~$0.958   $~ \\ \hline \end{tabular}
\caption{The wavefunction parameters which are fixed by using the method
of least squares for the CELLO, the CLEO, the BABAR and the BELLE data,
respectively. }  \label{tpptff} \end{table}

\begin{figure}[htb] \centering
\includegraphics[width=0.5\textwidth]{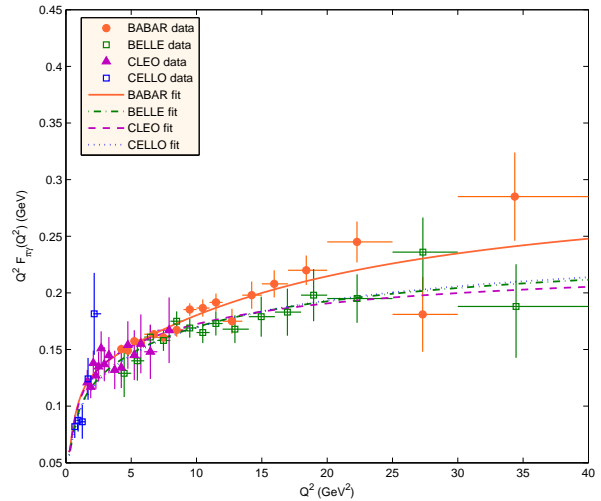} \caption{The pion-photon
TFFs $Q^2 F_{\pi\gamma}(Q^2)$ for the experimental data measured by the
BABAR, the BELLE, the CLEO and the CELLO Collaborations, respectively. The
fitted curves are obtained by using the method of least squares. }
\label{fpptff} \end{figure}

We adopt Eq.(\ref{pptff}) as the basic input function to achieve a best
fit of the pion wavefunction parameters by using the known experimental
data on the TFF $F_{\pi\gamma}(Q^2)$. More specifically, the values of the
two free parameters $(m_q, B)$ are fixed by requiring them to achieve the
minimum value of $\chi^2(m_q, B)$, which indicates a best fit of the
experimental data within the allowable parameter spaces. The determined
pion wavefunction parameters for the data of the BABAR, the BELLE, the
CLEO and the CELLO collaborations are presented in Table \ref{tpptff},
where the values of $\chi^2_{\rm min}/n_d$ and the probability
$P_{\chi^2_{\rm min}}$ are also presented. The pion-photon TFFs under
those parameters are put in Fig.(\ref{fpptff}). Fig.(\ref{fpptff}) shows
that the BELLE, the CLEO and the CELLO data result in similar trend of the
pion-photon TFF, while the BABAR data leads to a quite different TFF
behavior in larger $Q^2$-region, i.e. $Q^2 > 10\;{\rm GeV}^2$.

Table \ref{tpptff} shows a better fit with better confidence level can be
achieved from the CLEO and the BELLE data, whose probabilities are $0.986$
and $0.958$, respectively. The low probability of the CELLO data is
reasonable due to small number of data. The probability of the BABAR data
is less than $0.50$, indicating there may have some questionable points.
This conclusion agrees with the arguments of Refs.\cite{TFF_BB12,
TFF_BB142, TFF_BB134, TFF_BB14}. By using the EKHARA event
generator~\cite{EKHARA}, a Mont Carlo simulation of the pion-photon TFF on
the BESIII platform within the energy region $Q^2 < 3.1 {\rm GeV}^2$ has
been given in Ref.\cite{BESIII}. Those simulation data lead to: $m_q =
272$ MeV, $B = 0.058$, $A=22.118 {\rm GeV}^{-1}$, $\beta = 0.656$ GeV,
$\chi^2_{\rm min}/n_d = 4.521/16$ and $P_{\chi^2_{\rm min}} = 0.998$,
which is also consistent with the above BELLE, CELLO and CLEO
predictions.

\section{The pion wavefunction from the BELLE and the CLEO data}

In the above section, the pion wavefunction parameters are fixed by
minimizing the likelihood function $\chi^2$. In present section, we shall
adopt a weaker constraint from the probability $P_{\chi^2}$ to do a more
detailed discussion on possible constraints on the pion wavefunction. This
is reasonable, since the present data themselves are of certain
uncertainties and we do not need to require the theoretical prediction to
fit the data extremely well. The future more precise data shall lead to
more strict constraints. In this section, we shall only adopt the BELLE
and CLEO data to do the discussion, since they are at the more confidence
level.

\begin{figure}[htb] \centering
\includegraphics[width=0.5\textwidth]{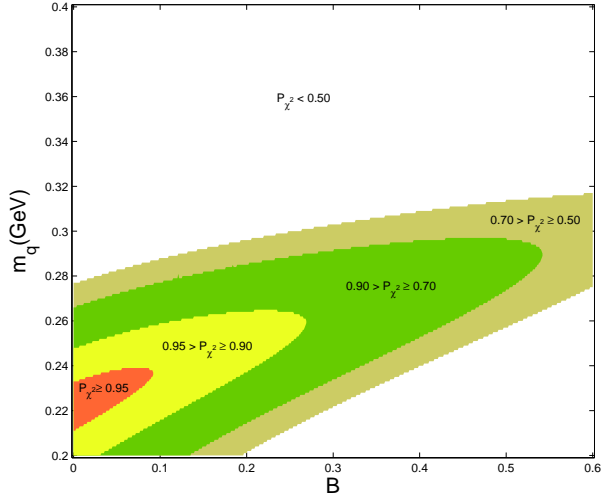} \caption{The allowable
$(m_q,B)$-region versus the probability $P_{\chi^2}$ from the BELLE data
for $Q^2\in[4,40]\rm GeV^2$. The four shaded bands from inside to outside
are for $P_{\chi^2}\geq 95\%$, $90\% \leq P_{\chi^2} < 95\%$, $70\% \leq
P_{\chi^2} < 90\%$ and $50\%\leq P_{\chi^2}< 70\%$, respectively. }
\label{fmBBELLE} \end{figure}

\begin{figure}[htb] \centering
\includegraphics[width=0.5\textwidth]{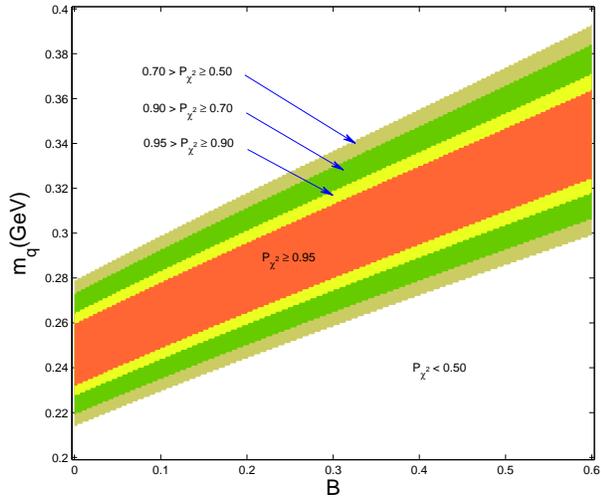} \caption{The allowable
$(m_q,B)$-region versus the probability $P_{\chi^2}$ from the CLEO data
for $Q^2 \in [1.5,9.2]{\rm GeV}^2$. The four shaded bands from inside to
outside are for $P_{\chi^2}\geq 95\%$, $90\% \leq P_{\chi^2} < 95\%$,
$70\% \leq P_{\chi^2} < 90\%$ and $50\%\leq P_{\chi^2}< 70\%$,
respectively.}  \label{fmBCLEO} \end{figure}

Figs.(\ref{fmBBELLE},\ref{fmBCLEO}) show the allowable $(m_q,B)$-region
versus the probability $P_{\chi^2}$ from either the BELLE or CLEO data,
where the four shaded bands from inside to outside are for $P_{\chi^2}\geq
95\%$, $90\% \leq P_{\chi^2} < 95\%$, $70\% \leq P_{\chi^2} < 90\%$ and
$50\%\leq P_{\chi^2}< 70\%$, respectively.
Figs.(\ref{fmBBELLE},\ref{fmBCLEO}) show that a more strict constraint to
the parameters can be achieved by a more bigger probability $P_{\chi^2}$.

\begin{figure}[htb] \centering
\includegraphics[width=0.5\textwidth]{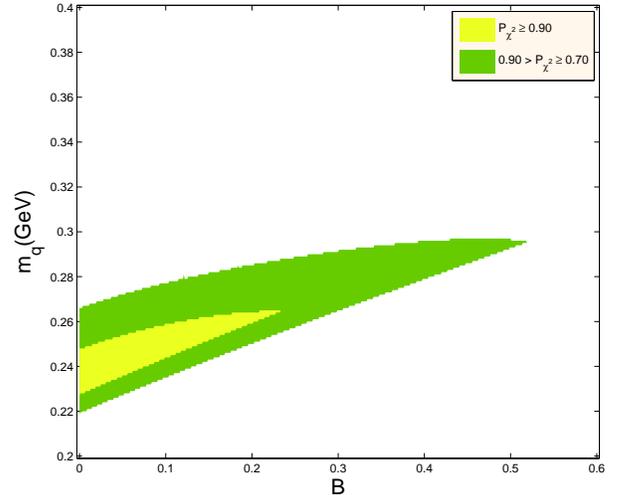} \caption{The
allowable $(m_q,B)$-region from both the BELLE and the CLEO data, where
the light-color band is for $P_{\chi^2} \geq 90\%$ and the fuscous band is
for $70\% \leq P_{\chi^2} < 90\%$.} \label{fmbBELLECLEO} \end{figure}

As a combination, we put the allowable $(m_q,B)$-region from both the
BELLE and CLEO data in Fig.(\ref{fmbBELLECLEO}). Fig.(\ref{fmbBELLECLEO})
shows that if requiring $P_{\chi^2}\geq 90\%$, we shall obtain $B \in
[0,0.235]$ and $m_q \in [227,265] {\rm MeV}$, which then lead to the range
of the first two Gegenbauer moments of the pion DA: $a_2(1{\rm GeV}) =
[0.087,0.348]$ and $a_4(1{\rm GeV}) = [-0.007,0.015]$. The predicted $a_2$
agrees with the ones determined in the literature from other approaches or
other processes, i.e. $a_2(1{\rm GeV}) =
0.26^{+0.21}_{-0.09}$~\cite{a2_SR04} and $0.19 \pm 0.06$~\cite{a2_SR98} by
QCD sum rules on the pion-photon TFFs; $0.24 \pm 0.14 \pm
0.08$~\cite{a2_LCSR00}, $0.20 \pm 0.03$~\cite{a2_LCSR05} and $0.19 \pm
0.05$~\cite{a2_LCSR002} by QCD LCSRs on the pion form factors; $0.19 \pm
0.19 \pm 0.08$~\cite{a2_LCSR052}, $0.17^{+0.15}_{-0.17}$~\cite{a2_LCSR053}
and $0.112 \pm 0.073$~\cite{a2_LCSR13} by LCSRs analysis on the $B/D \to
\pi l\nu$.

It is noted that $a_2 \sim B$, indicating that the longitudinal behavior
of the pion wavefunction is dominantly determined by the parameter $B$.
The BELLE data provides a strong constraint for both $B$ and $m_q$,
especially for $P_{\chi^2}\geq90\%$. On the other hand,
Fig.(\ref{fmBCLEO}) shows by using the lower $Q^2$-data alone, one cannot
determine the pion wavefunction's longitudinal behavior, because in the
low $Q^2$-region, the TFF is insensitive to the choice of the parameter
$B$ \footnote{For a bigger $B$, one only needs a reasonable bigger
constitute quark mass $m_q$ to get the same TFF. This observation agrees
with the prediction of Ref.\cite{TFF_NV}.}. However as will be shown
later, the low-energy data is helpful for determining the transverse
behavior of the pion wavefunction.

\begin{figure}[htb] \centering
\includegraphics[width=0.5\textwidth]{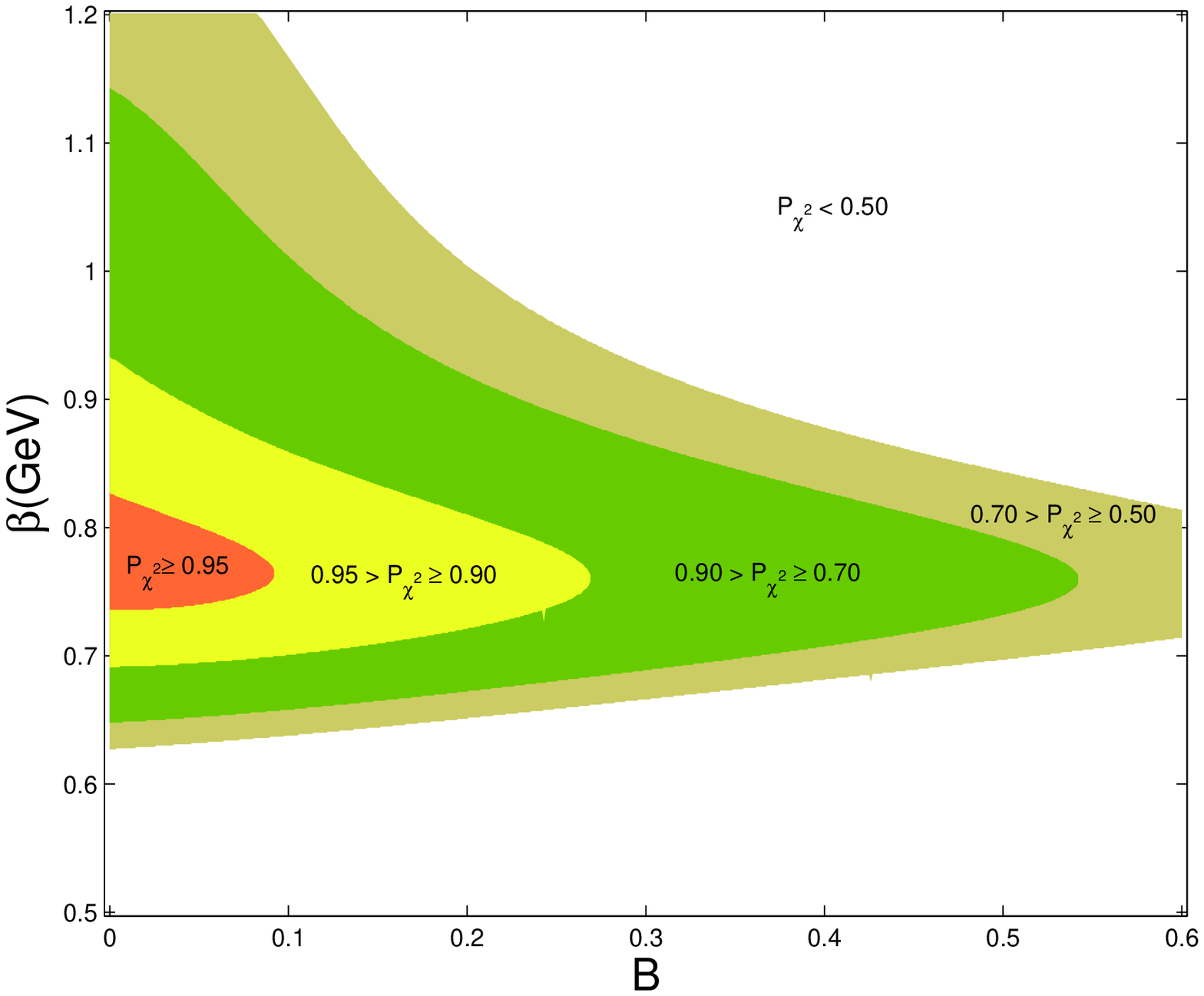} \caption{The
allowable $(\beta,B)$-region versus the probability $P_{\chi^2}$ from the
BELLE data, where $Q^2 \in [4,40]\rm GeV^2$. The four shaded bands from
inside to outside are for $P_{\chi^2}\geq 95\%$, $90\% \leq P_{\chi^2} <
95\%$, $70\% \leq P_{\chi^2} < 90\%$ and $50\%\leq P_{\chi^2}< 70\%$,
respectively. } \label{fmbetaBELLE} \end{figure}

\begin{figure}[htb] \centering
\includegraphics[width=0.5\textwidth]{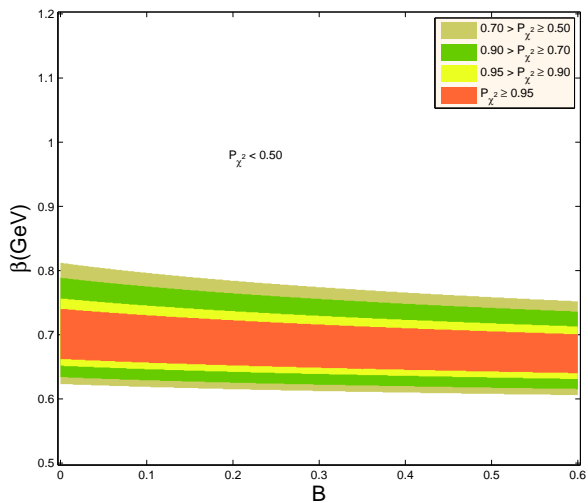} \caption{The
allowable $(\beta,B)$-region versus the probability $P_{\chi^2}$ from the
CLEO data, where $Q^2 \in [1.5,9.2]\rm GeV^2$. The four shaded bands from
inside to outside are for $P_{\chi^2}\geq 95\%$, $90\% \leq P_{\chi^2} <
95\%$, $70\% \leq P_{\chi^2} < 90\%$ and $50\%\leq P_{\chi^2}< 70\%$,
respectively. } \label{fmbetaCLEO} \end{figure}

The transverse behavior of the pion wavefunction is dominated by the
harmonic parameter $\beta$~\cite{Huang:1994dy}. To show how the
experimental data affect the transverse behavior, we take the parameters
($B$,$\beta$) as the two free input parameters. Following the same fit
procedures, we can obtain the allowable ranges for the parameters
($B$,$\beta$). The results are presented in
Figs.(\ref{fmbetaBELLE},\ref{fmbetaCLEO}), which are for the BELLE and the
CLEO data, respectively. Here the four shaded bands from inside to outside
are for $P_{\chi^2}\geq 95\%$, $90\% \leq P_{\chi^2} < 95\%$, $70\% \leq
P_{\chi^2} < 90\%$ and $50\%\leq P_{\chi^2}< 70\%$, respectively.

The BELLE data leads to $\beta \in [0.691,0.933] {\rm GeV}$ and $B \in
[0.00,0.269]$ for $P_{\chi^2} \geq 90\%$. Fig.(\ref{fmbetaBELLE}) shows
the allowed $\beta$ range shall be quickly expanded when $P_{\chi^2}$
becomes smaller. For example, when $B=0.20$, we have
$\beta\in[0.721,0.810]$ for $P_{\chi^2} \geq 90\%$,
$\beta\in[0.672,0.918]$ for $P_{\chi^2} \geq 70\%$ and
$\beta\in[0.651,1.004]$ for $P_{\chi^2} \geq 50\%$. On the other hand, as
shown by Fig.(\ref{fmbetaCLEO}), the low-energy CLEO data can give a
better constraint to the range of $\beta$, whose allowable range slightly
increases with the decrement of $P_{\chi^2}$. For example, we obtain
$\beta\in[0.652,0.757] {\rm GeV}$ for $P_{\chi^2} \geq 90\%$,
$\beta\in[0.634,0.789] {\rm GeV}$ for $P_{\chi^2} \geq 70\%$ and
$\beta\in[0.623,0.812] {\rm GeV}$ for $P_{\chi^2} \geq 50\%$,
accordingly.

\begin{figure}[htb] \centering
\includegraphics[width=0.5\textwidth]{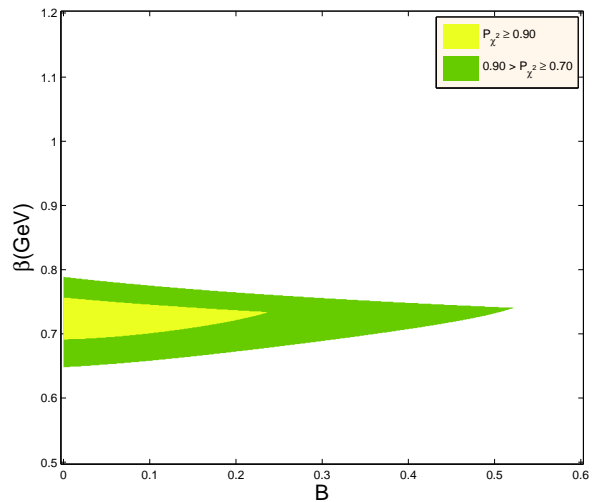} \caption{The
regions of the pion wavefunction parameters $(\beta,B)$ from the BELLE and
CLEO data under the probability $P_{\chi^2} \geq 90\%$, where the
light-color band is for $P_{\chi^2} \geq 90\%$ and the fuscous band is for
$70\% \leq P_{\chi^2} < 90\%$.}  \label{fmbetaBELLECLEO} \end{figure}

The allowable $(\beta,B)$-region from both the BELLE and CLEO data is
presented in Fig.(\ref{fmbetaBELLECLEO}). The lower edge of the shaded
band is determined by the BELLE data and the upper edge of the shaded band
is determined by the CLEO data, which indicates that the low-energy data
is important and helpful for determining the pion wavefunction's
transverse behavior. Fig.(\ref{fmbetaBELLECLEO}) shows that $B \in
[0.00,0.235]$ and $\beta \in [0.691,0.757] {\rm GeV}$ for $P_{\chi^2} \geq
90\%$.

\begin{figure}[htb] \centering
\includegraphics[width=0.5\textwidth]{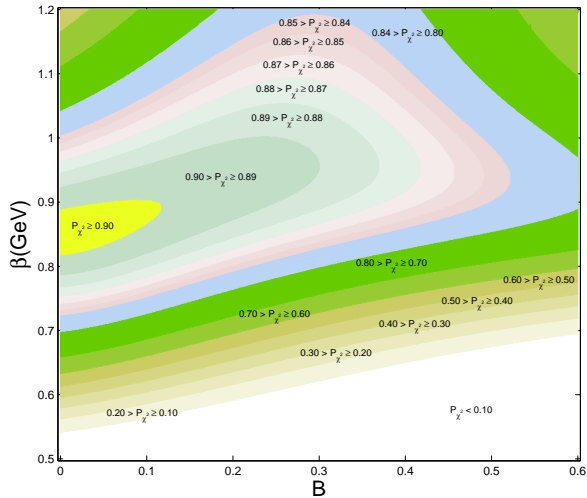} \caption{The
allowable $(\beta,B)$-region versus the choice of the probability
$P_{\chi^2}$ from the BELLE data in the high-energy region $Q^2\in[10,40]
{\rm GeV}^2$.} \label{fmbetaBELLEhigh} \end{figure}

The importance of the low-energy data can be further explained by
Fig.(\ref{fmbetaBELLEhigh}), which shows the allowable $(\beta,B)$-region
versus the probability $P_{\chi^2}$ from the BELLE data in high-energy
region $Q^2\in[10,40] {\rm GeV}^2$. Fig.(\ref{fmbetaBELLEhigh}) shows the
allowable range of $\beta$ is quickly broadened for a smaller and smaller
$P_{\chi^2}$. For example, when setting $B=0.00$, we shall have
$\beta\in[0.818,0.887]$ for $P_{\chi^2} \geq 90\%$,
$\beta\in[0.659,1.111]$ for $P_{\chi^2} \geq 70\%$ and
$\beta\in[0.613,1.2]$ for $P_{\chi^2} \geq 50\%$. Thus by using the large
$Q^2$ data alone, one may not get a definite conclusion on the transverse
behavior, unless the goodness-of-fit is high enough.

\section{Summary}

We have studied the transverse and longitudinal behavior of the pion
wavefunction by fitting the CELLO, the CLEO, the BABAR and the BELLE data
on the pion-photon TFF $F_{\pi\gamma}(Q^2)$. The method of least squares
is adopted for such an analysis.

\begin{figure}[htb] \centering
\includegraphics[width=0.5\textwidth]{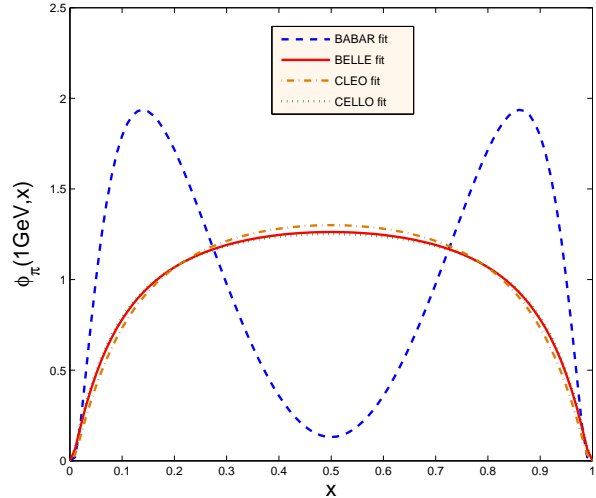} \caption{The pion DAs with
the parameters listed in Table \ref{tpptff}, which are fit from the TFF
data of BABAR, BELLE, CLEO and CELLO collaborations, respectively. }
\label{fphi} \end{figure}

Using the best fit parameters that lead to minimized likelihood function,
which are listed in Table \ref{tpptff}, we can get useful information on
the pion wavefunction. As an example, we put its distribution amplitude in
Fig.(\ref{fphi}). It is shown that the best fit of the CELLO, the CLEO and
the BELLE data prefer asymptotic-like behavior, while the BABAR data
prefers the more broad distribution, such as the CZ-like behavior. Table
\ref{tpptff} also indicates that a better fit with better confidence level can be achieved from the CLEO and the BELLE data, whose probabilities are
close to $1$. The low probability of the CELLO data is reasonable due to
small number of data. The probability of the BABAR data is less than
$0.50$, indicating there may have some questionable points within the
measured data.

The transverse and longitudinal behavior of the pion wavefunction is
dominantly determined by the parameter $\beta$ and $B$, respectively. The
parameter $B$ can be constrained by $Q^2 F_{\pi\gamma}(Q^2)$ behaviors in
high energy region precisely. For example, the BELLE data can determine
the parameter $B$ well. Figs.(\ref{fmBCLEO},\ref{fmbetaCLEO}) show that if
using the lower $Q^2$-data alone, such as the CLEO data, one cannot
determine the pion wavefunction's longitudinal behavior. However, as shown
by Fig.(\ref{fmbetaCLEO}), the low-energy CLEO data is important and
helpful for determining the transverse behavior of the pion wavefunction.
However, one still cannot determine the pion wavefunction precisely due
to the dramatic difference between the BABAR and BELLE data in the large
$Q^2$-region. Therefore, the future experimental data in the large
$Q^2$-region will be crucial for determining the longitudinal behavior of
the pion wavefunction.

\begin{figure}[htb]
\centering
\includegraphics[width=0.5\textwidth]{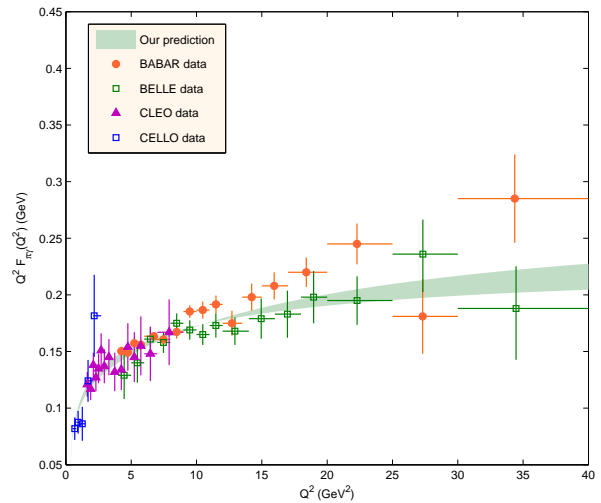}
\caption{The
predicted pion-photon TFF $Q^2 F_{\pi\gamma}(Q^2)$ by using the parameters
determined from the BELLE and the CLEO data. The BABAR, the BELLE, the
CLEO and the CELLO data are also presented as a comparison. }
\label{fpptffBELLECLEO}
\end{figure}

Using the BELLE and the CLEO data together and requiring $P_{\chi^2}\geq
90\%$, we obtain, $B \in [0,0.235]$, $m_q \in [227,265] {\rm MeV}$ and
$\beta \in [0.691,0.757] {\rm GeV}$. Using those parameters, our final
prediction on the pion-photon TFF $F_{\pi\gamma}(Q^2)$ are presented in
Fig.(\ref{fpptffBELLECLEO}).

{\bf Acknowledgments}: This work was supported in part by the Natural
Science Foundation of China under Grants No.11235005 and No.11275280, and
by the Fundamental Research Funds for the Central Universities under Grant
No.CDJZR305513.

\end{document}